\documentclass[preprint,floatfix]{revtex4}
\usepackage{graphicx}
\usepackage{amsmath}

\begin{document}
\title{Fixed-gap ground-state quantum computation and quantum concurrent processing}
\author{Ari Mizel}
\affiliation{Laboratory for Physical Sciences, College Park, Maryland 20740, USA}
\date{\today} 

\begin{abstract}
Quantum computation has revolutionary potential for speeding algorithms and for simulating quantum systems such as molecules.  We report here a quantum computer design that performs universal quantum computation within a single non-degenerate ground state protected from decohering noise by an energy gap that we argue is {\it system-size-independent}.  Closely analogous to a traditional electric circuit, it substantially changes the requirements for quantum computer construction, easing measurement, timing, and heating problems.  Using the standard adiabatic condition, we present evidence that this design permits ``quantum concurrent processing" distributing a quantum computation among extra qubits to perform a  quantum algorithm of $N$ gates in an amount of time that scales with $\sqrt{N}$.  One consequence of our work is a fixed gap version of adiabatic quantum computation, which several arguments hinted could be impossible.
\end{abstract}

\maketitle

The standard gate model \cite{Nielsen2000} for realizing a quantum computer involves fashioning a collection of two state quantum systems -- qubits -- that can be individually manipulated and measured.  By pulsing the Hamiltonian of each qubit in a precisely timed fashion, one subjects the collection of qubits to a desired succession of unitary gates.  If the experimental arrangement is capable of applying a sufficiently rich set of gates, universal computation is possible in which the initial state of the qubits $\left|\psi(t_0)\right>$ can be taken along a time-dependent trajectory $\left|\psi(t_1)\right>, \left|\psi(t_2)\right>, \dots$ to an arbitrary final state $\left|\psi(t_N)\right>$.

An alternate model, called ground-state quantum computation (GSQC) \cite{Mizel01,Mizel02,Mizel04,Mizel07}, replaces the pulsed, time-dependent state with a ``history'' state $\left|\Psi\right>$ in a larger Hilbert space \footnote{To the best of our knowledge, the apt term "history state" was coined by the authors of \cite{Aharonov04} to describe states of this type.  We sometimes find it helpful to describe the history state as quantum graph paper on which the time-dependent trajectory is recorded: $\left|\psi(t)\right>$ versus $t$.}.  This history state possesses a time-independent record of the entire evolution of the time-dependent state, so that $\left|\Psi\right>$ contains all of the amplitudes in $\left|\psi(t_0)\right>$, $\left|\psi(t_1)\right>, \dots, \left|\psi(t_N)\right>$.    

\begin{figure*}
[ptb]
\begin{center}
\includegraphics[
height=3.8207in
]%
{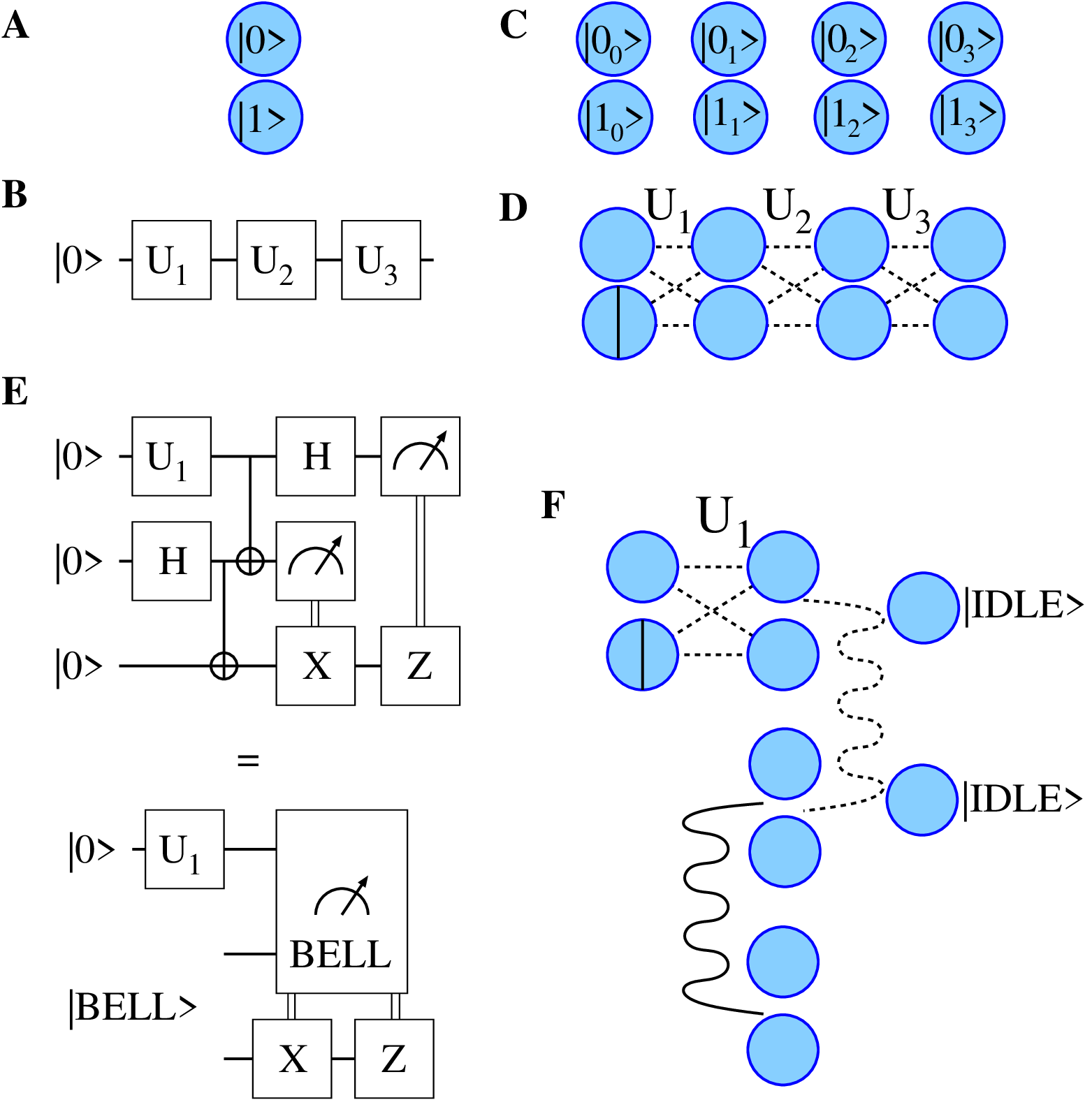}%
\caption{(A) Schematic of Hilbert space of a qubit.  (B) Standard gate model depiction of a single qubit undergoing an $N=3$ step calculation.  (C) GSQC Hilbert space of a single qubit undergoing an $N=3$ step calculation.  (D) Hamiltonian for calculation in (C).  Solid black line represents one-body onsite perturbation and dashed lines represent one-body tunneling energies.  (E) Standard gate model depiction of quantum teleportation with simplified representation.  (F) GSQC version of simplified representation of quantum teleportation.  Solid wavy line represents two-body onsite interaction and dashed wavy line represents two-body tunneling.}%
\label{fig:teleportcircuit}%
\end{center}
\end{figure*}

GSQC works by establishing a Hamiltonian whose ground state is the history state.  Let us start with the case of a single qubit possessing two quantum states $\left|0\right>$ and $\left|1\right>$ (see Fig.~\ref{fig:teleportcircuit}(A)).  Suppose that a quantum algorithm specifies that our qubit begin at stage $i=0$ of the computation in the initial state $\left|0\right>$ and undergo $N$ single qubit gates, given by the $2 \times 2$ unitary operators $U_i$, $i=1,...,N$ (see Fig.~\ref{fig:teleportcircuit}(B) for $N=3$).  The state at stage $i$ is thus supposed to be $\left|\psi(t_i)\right> = U_iU_{i-1} \dots U_1\left|0\right>$.  We wish to capture in a ground state the history of this single qubit through all $N+1$ computational stages.  So, we define a Hilbert space with $2 (N+1)$ states labelled by $\left|b_s\right>$, where $b=0,1$ is the bit value and $s=0,...,N$ is the computational stage (see Fig.~\ref{fig:teleportcircuit}(D)).  Our history state is $\left|\Psi \right> =  \sum_{i=0,...,N} U_i \dots U_1 \left|0_i\right>/\sqrt{N+1}$, a superposition of the system state at all $N+1$ stages of the computation.  It is useful to define in the $2(N+1)$ dimensional Hilbert space a unitary operator ${\mathcal U}_i$ by $\left<b^\prime_{s^\prime}\right|{\mathcal U}_i\left|b_s\right> = (U_i)_{b^\prime,b} \delta_{s,s^\prime}$.  Note that ${\mathcal U}_i$ acts on the bit value $b$ of $\left|b_s\right>$ but does not change the stage variable $s$.  To advance the stage variable, we define operators $A_{s^\prime,s} = \sum_{b=0,1} \left|b_{s^\prime}\right>\left<b_s\right|$ that do not change the bit value $b$.  The identity operator is $I$.  Our history state can then be written $\left|\Psi \right> = (I + A_{N,N-1} {\mathcal U}_N) (I + A_{N-1,N-2} {\mathcal U}_{N-1}) ...(I + A_{1,0} {\mathcal U}_1) \left|0_0\right>/\sqrt{N+1}$.  The state $\left|\Psi \right>$ is the ground state of the Hamiltonian ${\mathcal H}/\epsilon = \left|1_0\right>\left<1_0\right| + \sum_{b,s} (I-A_{s,s-1} {\mathcal U}_s)\left| b_{s-1}\right> \left<b_{s-1} \right|(I-{\mathcal U}_s^\dagger A^\dagger_{s,s-1})$ where $\epsilon$ is the fixed energy scale of the Hamiltonian.  This Hamiltonian is depicted schematically in Fig.~\ref{fig:teleportcircuit}(D) where a solid black line represents the on-site perturbation $\epsilon \left|1_0\right>\left<1_0\right|$ and the dotted lines represent tunneling energies.  There is an energy barrier to the excited states that scales like $\epsilon/(N+1)^2$ \cite{Mizel02}.

Compared to the standard approach, GSQC requires a larger Hilbert space but demands much less time-dependent control of a quantum system and provides an energy barrier to decoherence \cite{Mizel01} that defends against noise without active intervention.  This is a fundamental improvement over active quantum error correction schemes that constantly manage the qubits, detecting and correcting errors as they arise or echoing the qubits to average errors away.  To see the value of this approach, note that with the exception of actual spin-1/2 particles, most proposed realizations of quantum computers could not function without a protective energy gap separating two low energy qubit states from higher energy leakage states.  GSQC takes this well-tested idea further, performing the entire multiple qubit computation within a single, non-degenerate ground state that is protected from noise by an energy gap.  In addition, the time-dependent pulsing Hamiltonian is discarded, allowing for less coupling to the outside world and less difficulty with heating and timing.  Although departing from the standard gate model may seem unfamiliar initially, the GSQC history state is simply analogous to an ordinary electric circuit \cite{Mizel04}, with the input and output connected by a time-independent spatial network of wires.  The standard gate model quantum computer is actually more exotic, behaving like an abacus, with qubits shifting back-and-forth in time like abacus beads.   In other words, the standard gate model is like a mathematical calculation in which each the input to each step is overwritten by the output.  GSQC is like a mathematical calculation in which all steps are shown on the page, so that each step can be checked against the next.   We show here that not only can incorrect steps be penalized with an energy barrier but also that the history state makes it possible to perform some steps concurrently.

We need to change the GSQC formalism as presented so far, since it suffers from at least two deficiencies if we wish to perform calculations with a large number of stages $N$.  First, if $\left| \Psi\right>$ is measured, the probability that it will collapse to stage $s=N$, where the answer $U_N U_{N-1}...U_1\left|0_N\right>$ is stored, decreases with $N$ like $1/(N+1)$.  Second, the energy gap from the computationally meaningful ground state to the first excited state decreases with $N$, so that the energy barrier to decoherence becomes weaker and weaker.   

To solve both problems simultaneously, we observe that, from the standpoint of the final stage subspace $\left\{\left|0_N\right>,\left|1_N\right>\right\}$, the probability $N/(N+1)$ that the history state is inhabiting another stage can be regarded as a sort of leakage.  It is known that quantum teleportation \cite{Bennett93,Gottesman99}, shown in Fig.~\ref{fig:teleportcircuit}(E), can be an effective means of combating leakage \cite{Aliferis07}.  One can perform teleportation in a GSQC context  \cite{Mizel04}.  This does involve a complication since a quantum computer operator who performs a measurement in GSQC will collapse the history state \footnote{One could imitate measurement by coherently interacting ancilla qubits with data qubits, but this is likely to be inefficient.}; the Bell measurement step of teleportation therefore requires non-trivial adaptation.  Rather than simulating a Bell measurement with 4 possible outcomes, GSQC instead projects on to a singlet with only one possible outcome.  This works as follows  (see Fig.~\ref{fig:teleportcircuit}(F)).  Imagine we give our initial qubit a 4 state Hilbert space $\left|b_s\right>$ where $b, s = 0,1$ and prepare it in the state $(I+A_{1,0} {\mathcal U}_1)\left|0_0\right>/\sqrt{2}$, just as in the single qubit ground state quantum computation described above.  Two qubits, each given a 2 state Hilbert space $\left|b_0\right>$, $b=0,1$, are in a Bell pair $(\left|0_0\right>\left|1_0\right>-\left|1_0\right>\left|0_0\right>)/\sqrt{2}$.  The three qubit state in the $2 \otimes 2 \otimes 4$ Hilbert space is thus $\frac{1}{\sqrt{2}}(\left|0_0\right>\left|1_0\right>-\left|1_0\right>\left|0_0\right>)\frac{1}{\sqrt{2}} (I + A_{1,0}{\mathcal U}_1)\left| 0_0\right>$.  It is the ground state of the static Hamiltonian ${\mathcal H} + {\mathcal H}_{Create \,\, pairs}$,
\[
{\mathcal H} = \epsilon I \otimes I \otimes \left|1_0\right>\left<1_0\right| + I \otimes I \otimes H({\mathcal U}_1),
\]
\[
{\mathcal H}_{Create \,\, pairs} = H_{Create \,\, pairs} \otimes I,
\]
\begin{equation}
H({\mathcal U})/\epsilon = (1-A_{1,0} {\mathcal U}) \sum_{b}\left| b_0 \right> \left<b_0\right|(1-{\mathcal U}^\dagger A_{1,0}^\dagger), \label{eq:H(U)}
\end{equation}
\begin{eqnarray}
H_{Create \,\, pairs}/\epsilon & = & \frac{1}{2} \left[(\left|0_0\right>\left|0_0\right>+\left|1_0\right>\left|1_0\right>)(\left<0_0\right|\left<0_0\right|+\left<1_0\right|\left<1_0\right|) \right. \nonumber \\
& & + (\left|1_0\right>\left|0_0\right>+\left|0_0\right>\left|1_0\right>)(\left<1_0\right|\left<0_0\right|+\left<0_0\right|\left<1_0\right|) \nonumber \\
& & \left. + (\left|0_0\right>\left|0_0\right>-\left|1_0\right>\left|1_0\right>)(\left<0_0\right|\left<0_0\right|-\left<1_0\right|\left<1_0\right|) \right]. \label{eq:Hcreatepairs}
\end{eqnarray}

We now incorporate the Bell projection which completes teleportation.    We will need to store the post-projection states of the projected qubits and therefore supplement the Hilbert space, supplying a 5th state $\left|\text{\text{IDLE}}\right>$ to the initial qubit and a 3rd state $\left|\text{IDLE}\right>$ to the right member of the Bell pair.  Thus, the net Hilbert space has dimension $2 \otimes 3 \otimes 5$ instead of $2 \otimes 2 \otimes 4$.  The (unnormalized) ground state, including the Bell state projection, is
\begin{equation}
\left|\Psi\right> = \frac{1}{\sqrt{2}}(\left|0_0\right>\left|1_0\right>-\left|1_0\right>\left|0_0\right>)\frac{1}{\sqrt{2}} (I + A_{1,0}{\mathcal U}_1)\left| 0_0\right> + \frac{\Lambda}{2\sqrt{2}} {\mathcal U}_1 \left|0_0\right> \left|\text{IDLE}\right>\left|\text{IDLE}\right>. \label{eq:1qubit1U}
\end{equation}
Here, $\Lambda$ is a real parameter that will be made large to increase the relative contribution of the post-teleportation state to $\left| \Psi\right>$.  The Hamiltonian of the system is ${\mathcal H} + {\mathcal H}_{Create \,\, Bell}$ where ${\mathcal H}_{Create \,\, Bell}$ is unchanged but now
${\mathcal H} = \epsilon I \otimes I \otimes \left|1_0\right>\left<1_0\right| + I \otimes H({\mathcal U}_1,\Lambda)$ where
\begin{eqnarray}
\lefteqn{H({\mathcal U},\Lambda)/ \epsilon = I \otimes H({\mathcal U})/\epsilon + \left|\text{IDLE}\right>\left<\text{IDLE}\right| \otimes \sum_{b,s = 0,1} \left|b_s\right>\left<b_s\right| + \sum_{b = 0,1} \left|b_0\right>\left<b_0\right| \otimes \left|\text{IDLE}\right>\left<\text{IDLE}\right|} \label{eq:H(U,lambda)} \\
& & + \frac{1}{1 + \Lambda^2} \left( \Lambda \frac{\left|0_0\right>\left|1_1\right>-\left|1_0\right>\left|0_1\right>}{\sqrt{2}} - \left|\text{IDLE}\right>\left|\text{IDLE}\right> \right) \left( \Lambda \frac{\left<0_0\right|\left<1_1\right|-\left<1_0\right|\left<0_1\right|}{\sqrt{2}}- \left<\text{IDLE}\right|\left<\text{IDLE}\right|\right). \nonumber 
\end{eqnarray}
Note that none of the matrix elements of Hamiltonian $H({\mathcal U},\Lambda)/\epsilon$ exceeds 1 irrespective of the value of $\Lambda$: there are no unphysical demands being made of our GSQC Hamiltonian.

Assessing our GSQC teleportation procedure, we note that originally a GSQC of a single gate ${\mathcal U}_1$ acting on a single qubit produced a state $(I + A_{1,0}{\mathcal U}_1)\left|0_0\right>/\sqrt{2}$.  Measurement extracted the answer ${\mathcal U}_1\left|0_1\right>$ at the final stage $s=1$ with probability $1/2$ and the result $\left|0_0\right>$ at the earlier stage $s=0$ with probability $1/2$.  Measurement of the leftmost qubit in the new $2 \otimes  3 \otimes 5$ dimensional Hilbert space now yields an answer at the final stage with certainty since this qubit has only 1 stage, but this answer is incorrect with probability $p = 6/(8+\Lambda^2)$.  We have mapped the GSQC leakage problem into the usual standard model problem of faulty gates.  Fortunately, this problem is well studied \cite{Shor96,Aharonov96,Knill98,Gottesman98,Preskill98} and one can eliminate the errors by applying a GSQC version of  quantum error detecting codes\footnote{Note that this approach is quite distinct from the idea in \cite{Jordan06} of using codes to make transitions to low energy states unlikely.  In our formulation, there are no low energy excited states; the codes are used to correct distortions of the ground state itself.  Our approach is also distinct from the work of \cite{Lidar08} that includes fast pulses; no fast time-dependence is involved in our method except for the final measurement at the end of the calculation to extract the answer from the quantum computer.}.  For instance, the GSQC version of the $[[4,1,2]]$ Bacon-Shor code \cite{Aliferis07} is described in the  Appendix.  Using the Hellman-Feynman theorem, one can argue that the energy gap to the first excited state satisfies the lower bound
\begin{equation}
\frac{E(\Lambda)}{E(\Lambda = 0)} \ge  \frac{e^{-2 \text{arcsinh}(\Lambda)}}{1+\Lambda^2} \sim \frac{1}{4 (1+\Lambda^2)^2} \sim  \left(\frac{p}{12}\right)^2 \label{eq:gap}
\end{equation}
where $E(\Lambda = 0)= (3-\sqrt{5})\epsilon /2$ (see Appendix).  This lower bound should apply to an arbitrary number of GSQC qubits undergoing an arbitrary number of GSQC gates.  In fact, this lower bound may underestimate the real gap substantially; small simulations even suggest $E(\Lambda)/E(\Lambda = 0) \gtrsim p$ rather than $\gtrsim p^2$ \footnote{It would be appealing to find an uncertainty relation of the form $E(\Lambda)/E(\Lambda=0) (1/p) \gtrsim 1$, but our simulation results may suffer from finite size effects.}.  Note that the lower bound increases with error probability -- a small error probability per gate requires a small gap.  For a GSQC in the laboratory, the minimum achievable distortion $p$ of the ground state is constrained by $E(p)  \gg k_B T$ where $k_B T$ is the temperature in units of energy.  Thus, the usual quantum computing problem of reducing the gate error probability $p$ below a threshold $p_{\text{threshold}}$  has been exchanged for the potentially easier problem of having a sufficiently low temperature and/or high Hamiltonian energy scale $\epsilon$ to allow $E(p_{\text{threshold}})\gg k_B T$.  Note that, in addition to the inherent error probability $6/(8+\Lambda^2)$, imperfections in fabrication must be added to $p$ when determining if $p < p_{\text{threshold}}$.  Such time-independent errors can be minimized by testing and refining the apparatus and by compensating for known distortions using GSQC versions of decoherence free subspaces \cite{Zanardi97,Lidar98}, dynamical decoupling \cite{Viola99}, and the like, while both time-dependent and time-independent errors will be suppressed by the energy gap (see Appendix).  
\begin{figure*}
[ptb]
\begin{center}
\includegraphics[
height=3.8207in
]%
{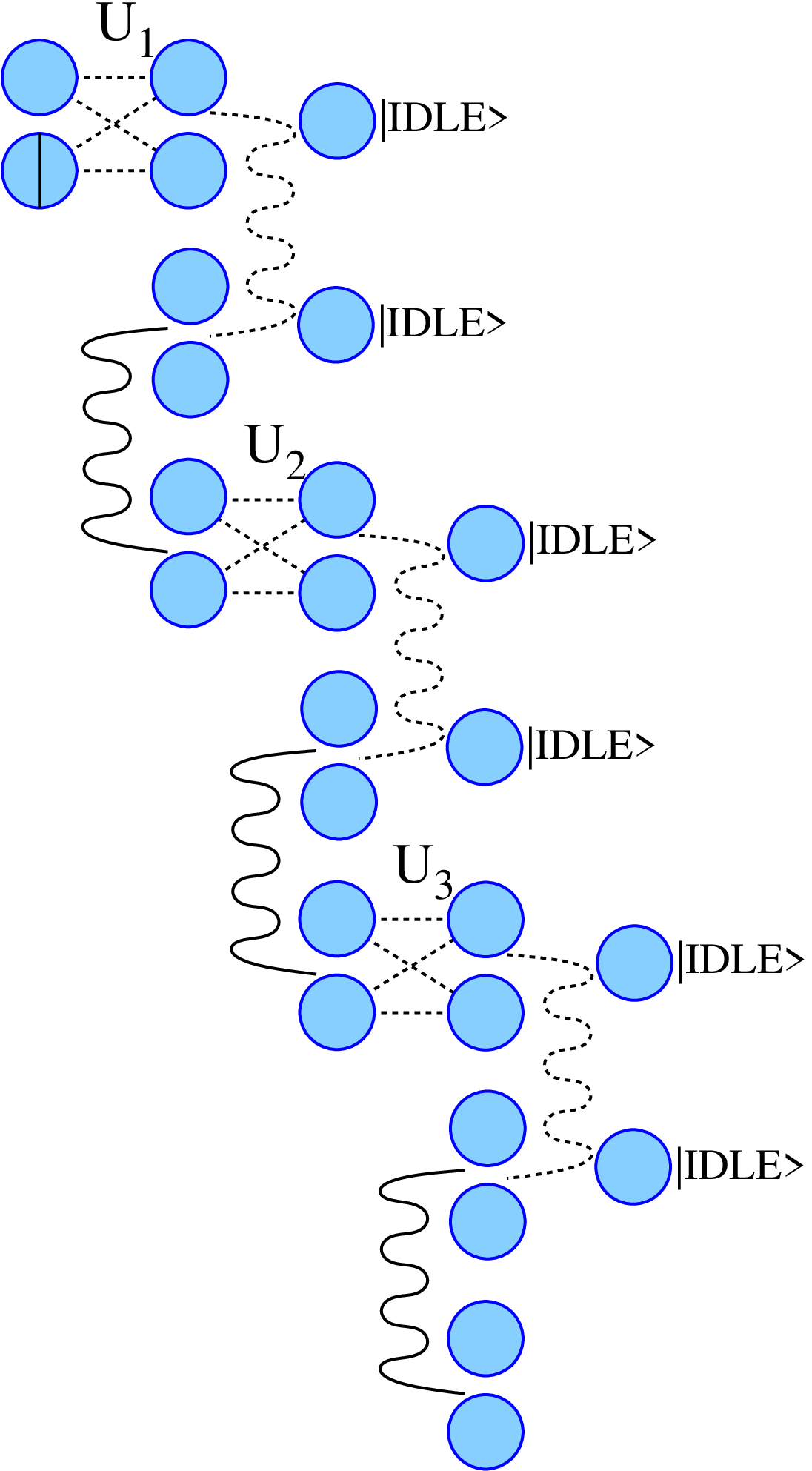}%
\caption{Iterated GSQC quantum teleportation that performs $N=3$ step calculation of Fig. ~\ref{fig:teleportcircuit}(D).}%
\label{fig:teleport_severalgates}%
\end{center}
\end{figure*}

It is straightforward to generalize to a single qubit calculation with $N$ operators ${\mathcal U}_1,\dots,{\mathcal U}_N$, as shown in Fig.~\ref{fig:teleport_severalgates} for the case $N=3$.  The Hilbert space is a tensor product $2 \otimes 3 \otimes 5 \otimes \dots \otimes 3 \otimes 5 \otimes 3 \dots \otimes 5$.    The generalization to multiple qubit calculations is similarly straightforward (Appendix).  This GSQC design requires an amount of hardware that scales linearly in the number of gates, like a traditional electric circuit with no clock cycle.

All of the steps of the calculation are simultaneously present in the GSQC history state, and GSQC appears to be capable of a new type of quantum speedup we term "quantum concurrent processing."  Our argument employs adiabatic quantum computation (AQC).  AQC and GSQC both involve quantum computation using ground states, but the theories have different origins.  In 2000, Farhi et al. suggested using adiabatic evolution to solve ``instances of the satisfiability problem" \cite{Farhi00,Farhi01}.  They continued to develop AQC as a creative methodology for attacking certain types of computational problems but not a form of universal computation that could execute existing algorithms.  In contrast, when it was proposed in 1999, GSQC had no new algorithmic content.  Instead, GSQC provided a universal history state and Hamiltonian suitable for computing, for instance, Shor's algorithm \cite{Shor94}, while reducing the need for time-dependent control and providing an energy gap for combating decoherence.  AQC gave an adiabatic recipe for reaching the ground state which they carefully distinguished from cooling \cite{Farhi00}; the adiabatic recipe can be justified using the adiabatic theorem.  GSQC, on the other hand, did imagine cooling to the ground state but never analyzed whether this would work for the GSQC Hamiltonian \footnote{A cooling process would not even require slow control of the Hamiltonian parameters, which would be a highly desirable experimental simplification and could avoid certain sources of decoherence, but the cooling time might be long.}.   After AQC was proposed, Mitchell realized it was probably straightforward to reach a GSQC ground state by adiabatically tuning the parameters of the GSQC Hamiltonian \footnote{See the 5th paragraph of \cite{Mizel02}, "In implementing a quantum ground-state computer, it is essential to address the issue of cooling time; one approach might be to turn on gradually the tunneling matrix elements in the Hamiltonian."}, albeit not necessarily in the canonical $H_0(1-s) + H_1 s$ fashion of AQC, and an explicit formulation of this ``squeegee'' method has appeared \cite{Mizel07}.  And since GSQC was proposed, AQC researchers have investigated the efficacy of an energy barrier to decoherence \cite{Childs02} and have come to embrace the idea of a universal history state \cite{Aharonov04}.

A number of recent developments have given the impression that the AQC energy gap shrinks  quickly with system size, leading to pessimism  \cite{Osborne06,Hastings09,Altshuler09} about AQC since a small gap translates into unacceptably long AQC calculation time and vulnerability to environmental excitations.  The version of GSQC formulated in this paper, however, yields a fixed-gap version of AQC as a corollary \footnote{The result in \cite{Hastings09} is irrelevant for us, because it concerns AQC in 1 dimension, whereas our GSQC design is 2 dimensional once multiple qubits are included.  The result in \cite{Altshuler09} pertains to Farhi et al.'s Hamiltonian of \cite{Farhi01}, not to a generic AQC Hamiltonian and in particular not to the one we present here.  Finally, \cite{Osborne06} assumes that the results of AQC should be stored in the expectation values of local operators, but this is not true for concatentated fault-tolerant quantum computing, in which the codes get larger as the number of steps in the algorithm grows.}.  To demonstrate fixed-gap AQC, one need only show that it is possible to tune the parameters of the fixed-gap GSQC Hamiltonian adiabatically to bring the system into the GSQC ground state.  This can be achieved by starting with $\Lambda = 0$ in the GSQC Hamiltonian.  The GSQC Hamiltonian is then extremely simple, essentially organizing the qubits into Bell pairs.  The gap is $(3-\sqrt{5})\epsilon/2$.  Next, we adiabatically take $\Lambda$ from $0$ to $\Lambda_{max}$ at a rate that depends on the system size.  The traditional adiabatic condition states that the system will remain in the ground state as long as the increase of $\Lambda$ occurs on a time 
\begin{equation}
T \gg \hbar \Lambda_{max} \left| \left<\Psi_k\right| d{\mathcal H}/d\Lambda \left| \Psi \right>\right|/E_{k}^2 \label{eq:adiabaticcondition}
\end{equation}
where $k$ is any excited eigenstate of the net Hamiltonian and we have simplified the denominator using the fact that the energy of $\left|\Psi\right>$ vanishes.  As is shown in the Appendix, we are guaranteed to satisfy this inequality if $T \gg \hbar \Lambda_{max} \sqrt{N \epsilon /E^{3}(\Lambda_{max})}$, where $E(\Lambda_{max})$ is given by (\ref{eq:gap}) and $N$ is the number of computational steps \footnote{It may be advantageous to increase the value of $\Lambda$ at different times for different gates in a multiple qubit algorithm; we have not yet studied this in detail.  See Appendix.}.

Reflecting on this finding, we see that it implies that GSQC allows the computation of an $N$ step quantum algorithm in $O(\sqrt{N})$ steps (not including a $O(\text{poly}(\text{log}(N)))$ factor for error correction).  For instance, consider a computation consisting of a series of $N$ black-box single qubit gates.  Since the gates are black boxes, we can devise no clever shortcuts to speed the algorithm; in the standard gate model, we would need $O(N)$ time to apply the gates in series.  Using GSQC, the result of this series can be obtained in a time $O(\sqrt{N})$ using $O(N)$ resources \footnote{The GSQC Hamiltonian requires $O(N)$ resources and one could employ $N$ scientists to set up the GSQC Hamiltonian in parallel in a time $O(1)$.}.  Since the calculation is distributed among $O(N)$ times as much hardware, we term this result "quantum concurrent processing."  It is quite distinct from the familiar "quantum parallelism" that requires clever algorithmic design; here there is no structure at all assumed about the algorithm.  It is also distinct from Grover's search of an unsorted database of $N$ items with $O(\sqrt{N})$ oracle calls -- there is no oracle involved in quantum concurrent processing.   Evidently, a degree of coordination of resources is possible in GSQC that is not possible classically; using entanglement, there is no need to wait for the first part of a task to be completed before the next part of the task gets started.

We gratefully acknowledge helpful comments from Marvin Kruger, Daniel Lidar, Keith Miller, Kevin Osborn, and Mark Wilde.

\pagebreak

\noindent {\bf Appendix}

\noindent {\em Universal quantum computation}

We have used first quantized notation in the main text to render our findings as accessible as possible, but this notation becomes unwieldy as the number of gates and qubits increases.  Therefore, we now return to the second quantized notation of  Refs. \cite{Mizel01,Mizel02,Mizel04,Mizel07}.  In this notation, the state $\left|b_s\right>$ of qubit $Q$ is expressed using a creation operator $c^\dagger_{Q,s,b}\left|\text{vac}\right>$ where $Q$ labels the qubit, $s$ labels the stage, and $b$ labels the bit value.  Grouping the creation operators into row vectors, we define $C^\dagger_{Q,s} \equiv \left[ c^\dagger_{Q,s,0} \,\,\, c^\dagger_{Q,s,1}\right]$.  For simplicity, we assume that the qubit particles have bosonic commutation relations, although this does not affect any results.  In this notation, the stage operator  $A_{s^\prime,s} =  \sum_{b=0,1} \left|b_{s^\prime}\right>\left<b_s\right|$ for qubit $Q$ just becomes $A_{Q,s^\prime,s} = C^\dagger_{Q,s^\prime} C_{Q,s}$.

Substituting the second quantized notation into the history state $\left|\Psi \right> = (I + A_{N,N-1} {\mathcal U}_N) (I + A_{N-1,N-2} {\mathcal U}_{N-1}) ...(I + A_{1,0} {\mathcal U}_1) \left|0_0\right>/\sqrt{N+1} = \sum_{i=0,...,N} U_i \dots U_1 \left|0_i\right>/\sqrt{N+1}$ for qubit $Q$ yields $\left|\Psi \right>_Q = (I + C^\dagger_{Q,N} U_{Q,N} C_{Q,N-1}) (I + C^\dagger_{Q,N-1}  U_{Q,N-1} C_{Q,N-2})  ...(I + C^\dagger_{Q,1} U_{Q,1} C_{Q,0}) C^{\dagger}_{Q,0} \left[\begin{array}{c} 1\\0\end{array}\right] \left|\text{vac}\right>/\sqrt{N+1}$.  The  Hamiltonian of this qubit
${\mathcal H}/\epsilon = \left|1_0\right>\left<1_0\right| + \sum_{b,s} (I-A_{s,s-1} {\mathcal U}_s)\left| b_{s-1}\right> \left<b_{s-1} \right|(I-{\mathcal U}_s^\dagger A^\dagger_{s,s-1})$
can be written as
 ${\mathcal H}^Q/\epsilon = C^\dagger_{Q,0} \left[ \begin{array}{cc} 0 & 0 \\ 0 & 1 \end{array}\right] C_{Q,0} + \sum_s (C^\dagger_{Q,s}-C^\dagger_{Q,s-1}U^{\dagger}_{Q,s})(C_{Q,s}- U_{Q,s}C_{Q,s-1})$. 

We now describe the GSQC teleportation process that applies $U_{Q,i}$ to qubit $Q$.  First, consider the case of a single gate $U_{Q,1}$, the case which was described in the main text and depicted in Fig.~\ref{fig:teleportcircuit}(F).    Label the three particles involved in the teleportation $Q_1$,$Q_2$, and $Q_3$.  Set the $\left|\text{IDLE}\right>$ states of $Q_1$ and $Q_2$ to $d^\dagger_{Q_1}\left| \text{vac}\right>$ and $d^\dagger_{Q_2}\left| \text{vac}\right>$ respectively.  The Hamiltonian is then ${\mathcal H}+ {\mathcal H}_{Create \,\, Bell}$ where

\begin{equation}
{\mathcal H} =  \epsilon C^\dagger_{Q_1,0} \left[ \begin{array}{cc} 0 & 0 \\ 0 & 1 \end{array}\right] C_{Q_1,0} + H(Q_1,Q_2,U_{Q,1},\Lambda)
\end{equation}
\begin{equation}
{\mathcal H}_{Create \,\, pairs} =  H_{Create \,\, pairs}(Q_2,Q_3)
\end{equation}
\begin{eqnarray}
\lefteqn{H(Q_1,Q_2,U_{Q,1},\Lambda)/\epsilon = (C^\dagger_{Q_1,1}-C^\dagger_{Q_1,0} U^{\dagger}_{Q,1})(C_{Q_1,1}- U_{Q,1}C_{Q_1,0})}  \nonumber \\
&&+ d^{\dagger}_{Q_2} d_{Q_2} (C^\dagger_{Q_1,0} C_{Q_1,0}+C^\dagger_{Q_1,1} C_{Q_1,1}) + C^\dagger_{Q_2,0}C_{Q_2,0} d^{\dagger}_{Q_1}d_{Q_1} \label{eq:H} \\
&& + \frac{1}{1 + \Lambda^2} \left( \Lambda \frac{ c_{Q_1,1,1}^{\dagger} c_{Q_2,0,0}^{\dagger}- c_{Q_1,1,0}^{\dagger} c_{Q_2,0,1} ^{\dagger}}{\sqrt{2}} - d^{\dagger}_{Q_1}d^{\dagger}_{Q_2} \right) \left( \Lambda \frac{c_{Q_2,0,0} c_{Q_1,1,1} - c_{Q_2,0,1} c_{Q_1,1,0}}{\sqrt{2}} - d_{Q_2}d_{Q_1}\right). \nonumber
\end{eqnarray}
and
\begin{eqnarray}
\lefteqn{H_{Create \,\, pairs}(Q_2,Q_3)/\epsilon =} \nonumber \\
& \frac{1}{2} & \left[(c^{\dagger}_{Q_3,0,0}c^{\dagger}_{Q_2,0,0}+c^{\dagger}_{Q_3,0,1}c^{\dagger}_{Q_2,0,1})(c_{Q_2,0,0}c_{Q_3,0,0}+c_{Q_2,0,1}c_{Q_3,0,1}) \right.  \nonumber \\
& & (c^{\dagger}_{Q_3,0,1}c^{\dagger}_{Q_2,0,0}+c^{\dagger}_{Q_3,0,0}c^{\dagger}_{Q_2,0,1})(c_{Q_2,0,0}c_{Q_3,0,1}+c_{Q_2,0,1}c_{Q_3,0,0}) \nonumber \\
& & \left. + (c^{\dagger}_{Q_3,0,0}c^{\dagger}_{Q_2,0,0}-c^{\dagger}_{Q_3,0,1}c^{\dagger}_{Q_2,0,1})(c_{Q_2,0,0}c_{Q_3,0,0}-c_{Q_2,0,1}c_{Q_3,0,1}) \right]. \label{eq:HCreatePairs}
\end{eqnarray}

The ground state of this Hamiltonian is 
\begin{eqnarray*}
\lefteqn{\left|\Psi\right> = \left[(I + \Lambda d^{\dagger}_{Q_2} d^{\dagger}_{Q_1}  \frac{c_{Q_2,0,0} c_{Q_1,1,1} - c_{Q_2,0,1} c_{Q_1,1,0}}{\sqrt{2}}) \frac{1}{\sqrt{2}} (I + C^{\dagger}_{Q_1,1} U_{Q,1}C_{Q_1,0})\right]}  \\
& & \hspace{2.5in} \frac{c^\dagger_{Q_3,0,0}c^\dagger_{Q_2,0,1}-c^\dagger_{Q_3,0,1}c^\dagger_{Q_2,0,0}}{\sqrt{2}} C^{\dagger}_{Q_1,0} \left[\begin{array}{c} 1\\0\end{array}\right] \left|\text{vac}\right> \\
& = & \left[ \frac{1}{\sqrt{2}}(c^\dagger_{Q_3,0,0}c^\dagger_{Q_2,0,1}-c^\dagger_{Q_3,0,1}c^\dagger_{Q_2,0,0})\frac{1}{\sqrt{2}} (I + C^{\dagger}_{Q_1,1} U_{Q,1}C_{Q_1,0}) C^{\dagger}_{Q_1,0} \left[\begin{array}{c} 1\\0\end{array}\right] \right. \\
& &\hspace{3.0in}  \left.  + \frac{\Lambda}{2\sqrt{2}}  C^{\dagger}_{Q_3,0} U_{Q,1}  \left[\begin{array}{c} 1\\0\end{array}\right] d^{\dagger}_{Q_2} d^{\dagger}_{Q_1} \right] \left|\text{vac}\right>.
\end{eqnarray*}

With the second quantized notation, it is straightforward to generalize to treat the case of a single qubit $Q$ undergoing a succession of gates $U_{Q,i}$, $i=1,...,N$ (shown in Fig.~\ref{fig:teleport_severalgates} for the case $N=3$).  It has Hamiltonian ${\mathcal H}+ {\mathcal H}_{Create \,\, Bell}$ where
\begin{equation}
{\mathcal H} =  \epsilon C^\dagger_{Q_1,0} \left[ \begin{array}{cc} 0 & 0 \\ 0 & 1 \end{array}\right] C_{Q_1,0} + \sum_{q=0}^{N-1} H(Q_{2q+1},Q_{2q+2},U_{Q,q+1},\Lambda)
\end{equation}
and 
\begin{equation}
{\mathcal H}_{Create \,\, pairs}  =  \sum_{q=0}^{N-1} H_{Create \,\, Bell}(Q_{2q+2},Q_{2q+3})
\end{equation}
in terms of (\ref{eq:H}) and (\ref{eq:HCreatePairs}).
 Its ground state has the form
 \begin{eqnarray}
\lefteqn{\left|\Psi\right> = \Pi_{q=0}^{N-1} \left[(I + \Lambda d^{\dagger}_{Q_{2q+2}} d^{\dagger}_{Q_{2q+1}}  \frac{c_{Q_{2q+2},0,0} c_{Q_{2q+1},1,1} - c_{Q_{2q+2},0,1} c_{Q_{2q+1},1,0}}{\sqrt{2}})\right. } \nonumber \\
&& \hspace{0.75in} \left. \frac{1}{\sqrt{2}} (I + C^{\dagger}_{Q_{2q+1},1} U_{Q,q+1}C_{Q_{2q+1},0})\right] \nonumber \\
& &\hspace{0.3in} \Pi_{q=0}^{N-1} \frac{c^\dagger_{Q_{2q+3},0,0}c^\dagger_{Q_{2q+2},0,1}-c^\dagger_{Q_{2q+3},0,1}c^\dagger_{Q_{2q+2},0,0}}{\sqrt{2}}C^{\dagger}_{Q_1,0} \left[\begin{array}{c} 1\\0\end{array}\right] \left|\text{vac}\right>. 
\end{eqnarray}

So far, we have addressed a single qubit $Q$ undergoing a series of single qubit gates $U_{Q,i}$ with teleportation steps in between the gates.  Of course, for two qubits $Q$ and $R$ each independently undergoing a series of single qubit gates $U_{Q,i}$ and $U_{R,j}$, the ground state would take the product form
 \begin{eqnarray}
\lefteqn{\left|\Psi\right> = \Pi_{q=0}^{N-1} \left[(I + \Lambda d^{\dagger}_{Q_{2q+2}} d^{\dagger}_{Q_{2q+1}}  \frac{c_{Q_{2q+2},0,0} c_{Q_{2q+1},1,1} - c_{Q_{2q+2},0,1} c_{Q_{2q+1},1,0}}{\sqrt{2}}) \frac{1}{\sqrt{2}} (I + C^{\dagger}_{Q_{2q+1},1} U_{Q,q+1}C_{Q_{2q+1},0})\right]} \nonumber \\
& & \hspace{0.3in} \Pi_{r=0}^{N-1} \left[(I + \Lambda d^{\dagger}_{R_{2r+2}} d^{\dagger}_{R_{2r+1}}  \frac{c_{R_{2r+2},0,0} c_{R_{2r+1},1,1} - c_{R_{2r+2},0,1} c_{R_{2r+1},1,0}}{\sqrt{2}}) \frac{1}{\sqrt{2}} (I + C^{\dagger}_{R_{2r+1},1} U_{R,r+1}C_{R_{2r+1},0})\right] \nonumber \\
& &\hspace{2.0in} \Pi_{q=0}^{N-1} \frac{c^\dagger_{Q_{2q+3},0,0}c^\dagger_{Q_{2q+2},0,1}-c^\dagger_{Q_{2q+3},0,1}c^\dagger_{Q_{2q+2},0,0}}{\sqrt{2}}C^{\dagger}_{Q_1,0} \left[\begin{array}{c} 1\\0\end{array}\right]  \nonumber \\
& &\hspace{2.0in} \Pi_{r=0}^{N-1} \frac{c^\dagger_{R_{2r+3},0,0}c^\dagger_{R_{2r+2},0,1}-c^\dagger_{R_{2r+3},0,1}c^\dagger_{Q_{2r+2},0,0}}{\sqrt{2}}C^{\dagger}_{R_1,0} \left[\begin{array}{c} 1\\0\end{array}\right] \left|\text{vac}\right>. \label{eq:Psi}
\end{eqnarray}
and the Hamiltonian would simply be the sum of
\begin{eqnarray*}
{\mathcal H} & = &  \epsilon C^\dagger_{Q_1,0} \left[ \begin{array}{cc} 0 & 0 \\ 0 & 1 \end{array}\right] C_{Q_1,0} + \sum_{q=0}^{N-1} H(Q_{2q+1},Q_{2q+2},U_{Q,q+1},\Lambda) \nonumber \\
& & +  \epsilon C^\dagger_{R_1,0} \left[ \begin{array}{cc} 0 & 0 \\ 0 & 1 \end{array}\right] C_{R_1,0} + \sum_{r=0}^{N-1}  H(R_{2r+1},R_{2r+2},U_{R,r+1},\Lambda)
\end{eqnarray*}
and 
\[
{\mathcal H}_{Create \,\, pairs}  =  \sum_{q=0}^{N-1} H_{Create \,\, Bell}(Q_{2q+2},Q_{2q+3}) + \sum_{r=0}^{N-1}  H_{Create \,\, Bell}(R_{2r+2},R_{2r+3}).
\]

A non-trivial quantum computation will require two qubit gates.  To include a controlled-PHASE gate between qubits $Q$ and $R$ at stage $s$, we  remove the $q=s-1$, $r=s-1$ factor in the product state (\ref{eq:Psi}), which is 
\begin{eqnarray*}
& & \left[(I + \Lambda d^{\dagger}_{Q_{2s}} d^{\dagger}_{Q_{2s-1}}  \frac{c_{Q_{2s},0,0} c_{Q_{2s-1},1,1} - c_{Q_{2s},0,1} c_{Q_{2s-1},1,0}}{\sqrt{2}}) \frac{1}{\sqrt{2}} (I + C^{\dagger}_{Q_{2s-1},1} U_{Q,s}C_{Q_{2s-1},0})\right] \nonumber \\
& & \hspace{0.25in}\left[(I + \Lambda d^{\dagger}_{R_{2s}} d^{\dagger}_{R_{2s-1}}  \frac{c_{R_{2s},0,0} c_{R_{2s-1},1,1} - c_{R_{2s},0,1} c_{R_{2s-1},1,0}}{\sqrt{2}}) \frac{1}{\sqrt{2}} (I + C^{\dagger}_{R_{2s-1},1} U_{R,s}C_{R_{2s-1},0})\right]. \nonumber
\end{eqnarray*}
We replace this factor with 
\begin{eqnarray*}
& & \left[(I + \Lambda d^{\dagger}_{Q_{2s}} d^{\dagger}_{Q_{2s-1}}  \frac{c_{Q_{2s},0,0} c_{Q_{2s-1},1,1} - c_{Q_{2s},0,1} c_{Q_{2s-1},1,0}}{\sqrt{2}}) \right. \nonumber \\
&& \left. (I + \Lambda d^{\dagger}_{R_{2s}} d^{\dagger}_{R_{2s-1}}  \frac{c_{R_{2s},0,0} c_{R_{2s-1},1,1} - c_{R_{2s},0,1} c_{R_{2s-1},1,0}}{\sqrt{2}})  \right. \nonumber \\
&&  \hspace{0.25in}  \left. \frac{1}{\sqrt{2}} (I + c^{\dagger}_{R_{2s-1},1,0}c_{R_{2s-1},0,0} C^{\dagger}_{Q_{2s-1},1}C_{Q_{2s-1},0} + c^{\dagger}_{R_{2s-1},1,1}c_{R_{2s-1},0,1} C^{\dagger}_{Q_{2s-1},1}\, Z \, C_{Q_{2s-1},0}) \right] \nonumber
\end{eqnarray*}
where $Z$ is the $2 \times 2$ Pauli matrix $\sigma_z$.  The quantum state is now no longer a product of the state of qubit $Q$ and the state of qubit $R$ since the controlled-PHASE gate entangles them.

\begin{figure*}
[ptb]
\begin{center}
\includegraphics[
height=3.8207in
]%
{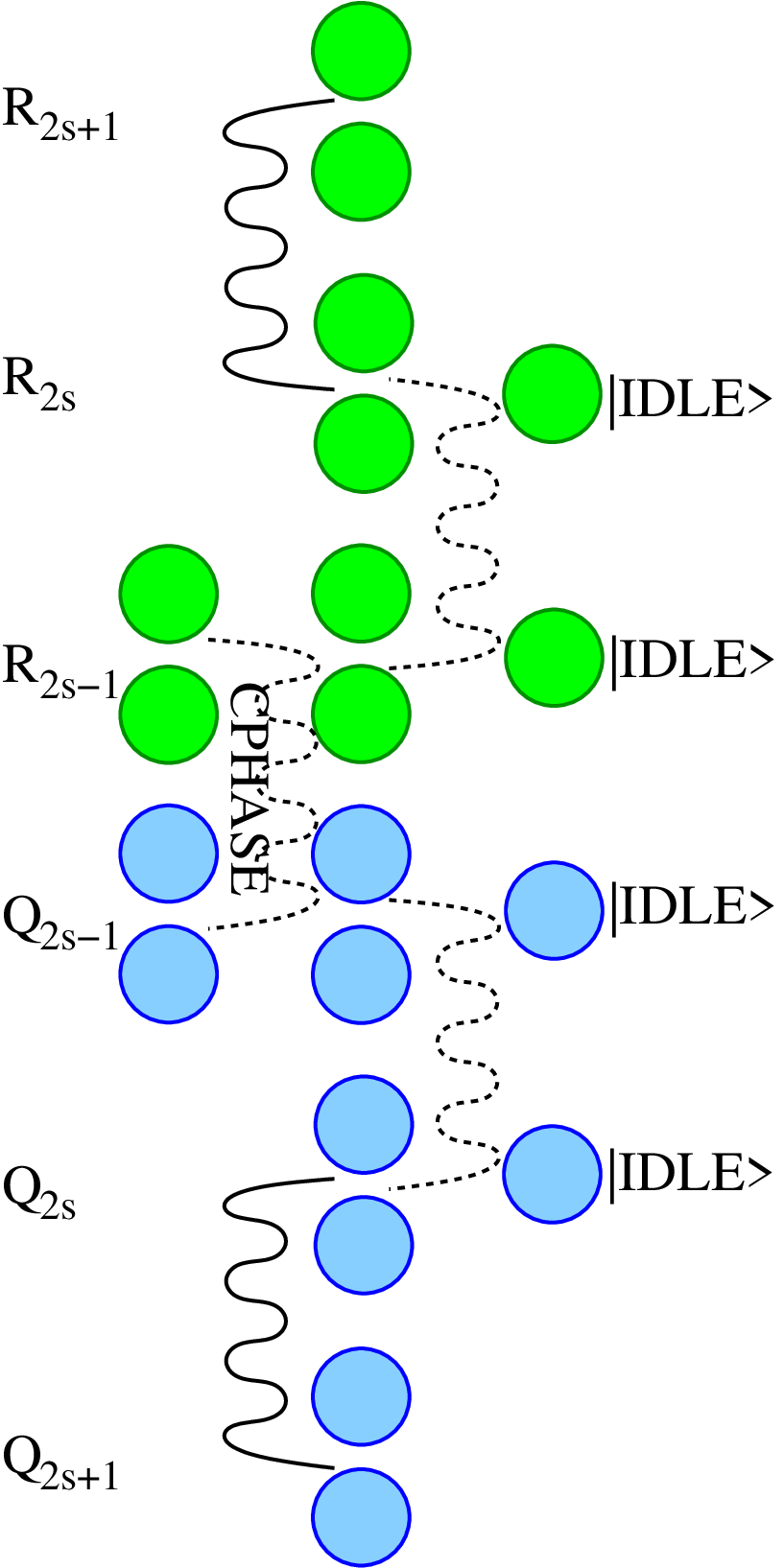}%
\caption{Schematic depiction of CPHASE gate.}%
\label{fig:teleport_CPHASE}%
\end{center}
\end{figure*}

To make this the ground state, we must alter the $q=s-1$ and $r=s-1$ terms of the Hamiltonian,
$H(Q_{2s-1},Q_{2s},U_{Q,s},\Lambda)+H(R_{2s-1},R_{2s},U_{R,s},\Lambda)$.  We replace the single qubit gate Hamiltonians $(C^\dagger_{Q_{2s-1},1}-C^\dagger_{Q_{2s-1},0} U^{\dagger}_{Q,s})(C_{Q_{2s-1},1}- U_{Q,s}C_{Q_{2s-1},0})$ and $(C^\dagger_{R_{2s-1},1}-C^\dagger_{R_{2s-1},0} U^{\dagger}_{R,s})(C_{R_{2s-1},1}- U_{R,s}C_{R_{2s-1},0})$ with a two qubit gate
\begin{eqnarray}
&& (c^{\dagger}_{R_{2s-1},1,1}C^\dagger_{Q_{2s-1},1}-c^{\dagger}_{R_{2s-1},0,1}C^\dagger_{Q_{2s-1},0} Z)(C_{Q_{2s-1},1} c_{R_{2s-1},1,1}- Z \,C_{Q_{2s-1},0}c_{R_{2s-1},0,1}) \nonumber \\
&& + (c^{\dagger}_{R_{2s-1},1,0}C^\dagger_{Q_{2s-1},1}-c^{\dagger}_{R_{2s-1},0,0}C^\dagger_{Q_{2s-1},0})(C_{Q_{2s-1},1} c_{R_{2s-1},1,0}- C_{Q_{2s-1},0}c_{R_{2s-1},0,0}) \nonumber \\
&& + (d^\dagger_{R_{2s-1}}d_{R_{2s-1}} + C^\dagger_{R_{2s-1},1}C_{R_{2s-1},1})C^\dagger_{Q_{2s-1},0}C_{Q_{2s-1},0} \nonumber \\
& & +(d^\dagger_{Q_{2s-1}}d_{Q_{2s-1}} + C^\dagger_{Q_{2s-1},1}C_{Q_{2s-1},1})C^\dagger_{R_{2s-1},0}C_{R_{2s-1},0}. \label{eq:replacementH}
\end{eqnarray}
The last line two lines impose an energy penalty if one of the qubits crosses through the gate without the other.  The CPHASE gate is depicted in Fig.~\ref{fig:teleport_CPHASE}.

Since we have considered both single qubit and CPHASE gates, we have a universal construction  \cite{Barenco95} for universal GSQC with teleportation.  Naturally, we could use a CNOT instead of a CPHASE by replacing the $2 \times 2$ $Z$ matrices in (\ref{eq:replacementH}) with $2\times 2$ $X$ matrices.  More generally, we can write
\begin{eqnarray*}
&& (C^{\dagger}_{R_{2s-1},1} \otimes C^\dagger_{Q_{2s-1},1}-C^{\dagger}_{R_{2s-1},0} \otimes C^\dagger_{Q_{2s-1},0} U^\dagger) (C_{R_{2s-1},1} \otimes C_{Q_{2s-1},1}-U C_{R_{2s-1},0} \otimes C_{Q_{2s-1},0}) \\
&& \hspace{1.0in} + (d^\dagger_{R_{2s-1}}d_{R_{2s-1}} + C^\dagger_{R_{2s-1},1}C_{R_{2s-1},1})C^\dagger_{Q_{2s-1},0}C_{Q_{2s-1},0}\\
& & \hspace{1.0in} +(d^\dagger_{Q_{2s-1}}d_{Q_{2s-1}} + C^\dagger_{Q_{2s-1},1}C_{Q_{2s-1},1})C^\dagger_{R_{2s-1},0}C_{R_{2s-1},0} 
\end{eqnarray*}
where $U$ is a $4 \times 4$ two qubit unitary matrix.  (In the CPHASE case, for instance, $U = U_{\text{CPHASE}} = \text{diag}(1,1,1,-1)$.)

\begin{figure*}
[ptb]
\begin{center}
\includegraphics[
height=3.8207in
]%
{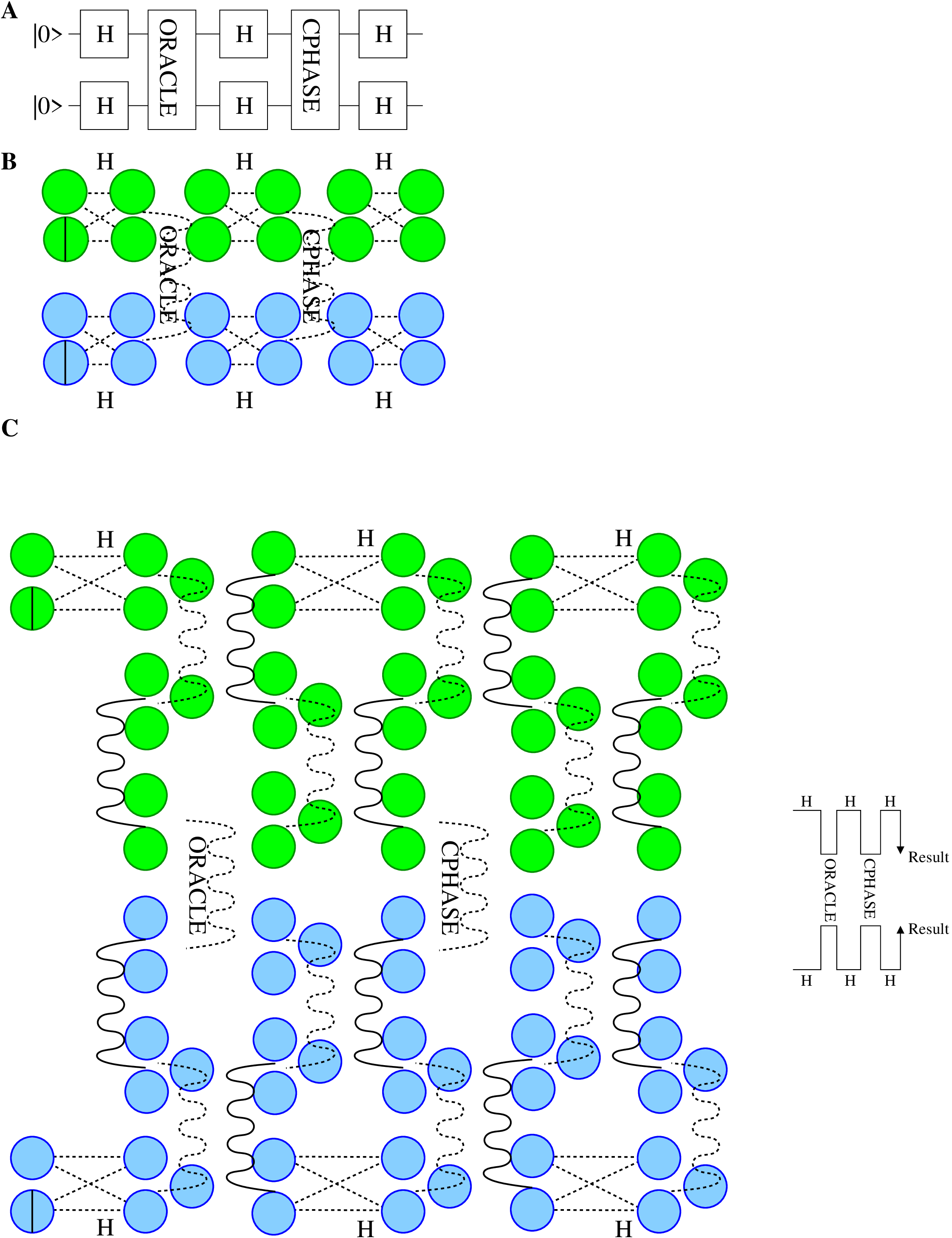}%
\caption{(A) Two qubit Grover algorithm in the standard gate model.  (B) Two qubit Grover algorithm in GSQC without teleportation.  This version of GSQC is not scalable since the gap decreases with circuit depth $N$.  (C) Two qubit Grover algorithm in fixed gap, teleportation version of GSQC.  Each qubit has a snaking geometry, as depicted in the small flow diagram.  Each gate requires a Hilbert space of size $2 \otimes 3 \otimes 5$ (e.g. 3 electrons in 10 dots).  The fabrication requirements are thus linear in the number of gates, just as in an ordinary electrical circuit.}%
\label{fig:Grover}%
\end{center}
\end{figure*}

A two qubit Grover \cite{Grover97} algorithm provides an illustrative example of a GSQC computation.  This algorithm has circuit depth $N=5$: the two qubits start in the product state $\left|0\right>\left|0\right>$, undergo Hadamard gates $H$, interact via an oracle (one of the 4 possible controlled phase gates), undergo Hadamard gates,  interact via the phase gate $I \otimes I  - 2 \left|00\right>\left<00\right|$, and undergo Hadamard gates.  The result is the computational basis state whose phase was flipped by the oracle.  The algorithm is shown in Fig.~\ref{fig:Grover} in standard gate model, GSQC, and teleportation-based GSQC versions.

\noindent {\em Energy Gap}

To compute a lower bound for the energy gap, we employ the Hellman-Feynman theorem.  This theorem states that for a parameter dependent Hamiltonian $H(\Lambda)$ with a non-degenerate eigenstate $\left|\Psi_k(\Lambda)\right>$ whose energy is $E_k(\Lambda)$,
\begin{equation}
\frac{d E_k}{d \Lambda} = \left< \Psi_k(\Lambda) \right| \frac{dH}{d\Lambda} \left | \Psi_k(\Lambda)\right>.
\end{equation}

Now, differentiating (\ref{eq:H}) gives
\begin{eqnarray*}
 \lefteqn{\frac{2}{(1 + \Lambda^2)^2} \left< \Psi_k \right| \left(  \frac{ c_{Q_1,1,1}^{\dagger} c_{Q_2,0,0}^{\dagger}- c_{Q_1,1,0}^{\dagger} c_{Q_2,0,1} ^{\dagger}}{\sqrt{2}} + \Lambda d^{\dagger}_{Q_1}d^{\dagger}_{Q_2} \right)} \nonumber \\
&&\hspace{1.0in}  \left( \Lambda \frac{c_{Q_2,0,0} c_{Q_1,1,1} - c_{Q_2,0,1} c_{Q_1,1,0}}{\sqrt{2}} - d_{Q_2}d_{Q_1}\right)\left| \Psi_k \right> \nonumber \\
= & &-\frac{2}{1 + \Lambda^2} \left< \Psi_k \right| \left(  \frac{ c_{Q_1,1,1}^{\dagger} c_{Q_2,0,0}^{\dagger}- c_{Q_1,1,0}^{\dagger} c_{Q_2,0,1} ^{\dagger}}{\sqrt{2}} + \Lambda d^{\dagger}_{Q_1}d^{\dagger}_{Q_2} \right) d_{Q_2}d_{Q_1} \nonumber \\
&&\frac{1}{1 + \Lambda^2} \left( \Lambda \frac{ c_{Q_1,1,1}^{\dagger} c_{Q_2,0,0}^{\dagger}- c_{Q_1,1,0}^{\dagger} c_{Q_2,0,1} ^{\dagger}}{\sqrt{2}} - d^{\dagger}_{Q_1}d^{\dagger}_{Q_2} \right) \left( \Lambda \frac{c_{Q_2,0,0} c_{Q_1,1,1} - c_{Q_2,0,1} c_{Q_1,1,0}}{\sqrt{2}} - d_{Q_2}d_{Q_1}\right)\left| \Psi_k \right> \nonumber \\
= & & -\frac{2}{1 + \Lambda^2} \left< \Psi_k \right| \left(  \frac{ c_{Q_1,1,1}^{\dagger} c_{Q_2,0,0}^{\dagger}- c_{Q_1,1,0}^{\dagger} c_{Q_2,0,1} ^{\dagger}}{\sqrt{2}} + \Lambda d^{\dagger}_{Q_1}d^{\dagger}_{Q_2} \right) d_{Q_2}d_{Q_1} (E_k-h(Q_1,Q_2))\left|\Psi_k\right> \nonumber
 \end{eqnarray*}
 where $h(Q_1,Q_2)$ is defined as the sum of terms in the Hamiltonian that do {\it not} involve $Q_1$ or $Q_2$ and where we have used the Schr\"odinger equation for $\left|\Psi_k\right>$.  Now,
 \begin{eqnarray*}
\lefteqn{-\frac{2}{1 + \Lambda^2} \left< \Psi_k \right| \left(  \frac{ c_{Q_1,1,1}^{\dagger} c_{Q_2,0,0}^{\dagger}- c_{Q_1,1,0}^{\dagger} c_{Q_2,0,1} ^{\dagger}}{\sqrt{2}} + \Lambda d^{\dagger}_{Q_1}d^{\dagger}_{Q_2} \right) d_{Q_2}d_{Q_1} (E-h(Q_1,Q_2))\left|\Psi_k\right>} \nonumber \\
=&&-\frac{1}{1 + \Lambda^2} \left< \Psi_k \right| \left[ \left(  \frac{ c_{Q_1,1,1}^{\dagger} c_{Q_2,0,0}^{\dagger}- c_{Q_1,1,0}^{\dagger} c_{Q_2,0,1} ^{\dagger}}{\sqrt{2}} + \Lambda d^{\dagger}_{Q_1}d^{\dagger}_{Q_2} \right) d_{Q_2}d_{Q_1} \right. \nonumber \\
& & \left. + d^\dagger_{Q_1}d^\dagger_{Q_2}\left(  \frac{ c_{Q_2,0,0}c_{Q_1,1,1}- c_{Q_2,0,1} c_{Q_1,1,0} }{\sqrt{2}} + \Lambda d_{Q_2}d_{Q_1} \right) \right] (E_k-h(Q_1,Q_2))\left|\Psi_k\right> \nonumber \\
\ge & & -\frac{\Lambda + \sqrt{1+\Lambda^2}}{1+\Lambda^2} \left< \Psi_k \right| (E_k-h(Q_1,Q_2))\left|\Psi_k\right>
 \end{eqnarray*}
since the smallest eigenvalue of the matrix $-\left[ \begin{smallmatrix} 0 & 1\\1 & 2\Lambda\end{smallmatrix} \right]$ is $-\Lambda-\sqrt{1+\Lambda^2}$.  Note that we have not proven this last inequality in all cases; this is why we use the term "argue" rather than the term "prove" in the main text and in the abstract.  Our final result is obtained by summing over $Q_i$ and $Q_j$
\begin{equation}
\frac{d E_k}{d \Lambda} \ge  -\frac{\Lambda + \sqrt{1+\Lambda^2}}{1+\Lambda^2}  \sum_{q=0}^{N-1} \left< \Psi_k \right| H - h(Q_{2q+1},Q_{2q+2}) \left | \Psi_k \right> \ge  -2 \frac{\Lambda + \sqrt{1+\Lambda^2}}{1+\Lambda^2} E_k 
\end{equation}
where the $2$ in the final equation arises because interaction terms between $Q_{2q+2}$ and $Q_{2q+3}$ appear in neither $h(Q_{2q+1},Q_{2q+2})$ nor $h(Q_{2q+3},Q_{2q+4})$.

This differential equation for $E_k(\Lambda)$ has the solution given in the main text,
\begin{equation}
\frac{E(\Lambda)}{E(\Lambda = 0)} \ge  \frac{e^{-2 \text{arcsinh}(\Lambda)}}{1+\Lambda^2}. 
\end{equation}
When $\Lambda = 0$, it is straightforward to obtain the eigenvalues of the Hamiltonian exactly.

\noindent {\em Error correcting codes in GSQC}

We established in the main text that GSQC allows fixed gap computation provided one allows a non-zero error probability per gate.  Fortunately, the problem of handling faulty gates is well studied in the context of the standard gate model; we need only adapt the conclusions of fault tolerant quantum error correction to GSQC.  Some complications do arise from the properties of measurement in GSQC.  One can only perform a measurement at the end of a GSQC, not in the middle of a calculation.  It is known that quantum error correction can be performed coherently without measurement, and this could be implemented in GSQC.  However, the noise threshold is expected to to decrease markedly if one refrains from measuring \cite{DiVincenzo07}.

Fortunately, as we showed in the main text in the case of Bell measurements, in GSQC one can modify the Hamiltonian to project \cite{Mizel04} a single result in the wavefunction rather than performing a genuine measurement with different possible outcomes.   This is particularly powerful in the case of error correction, since one can project the error syndrome measurement to a "no error" result.  (This in turn makes it possible to use error detecting codes rather than error correcting codes for GSQC if desired, since one always projects to a "no error" result and does not need to apply corrections.)

To see how this projection works, suppose that our data qubits have GSQC ground state $\left|\Psi\right>$ at some stage $s$ of the calculation.  We interact them with an ancilla to obtain the error syndrome, and, after the interaction, the zero-energy ground state of the Hamiltonian is $\left|0_0\right> \left|\Psi_0\right> + \left|1_0\right> \left|\Psi_1\right>$.  Here, the ancilla state $\left|0_0\right>$ indicates a "no error" syndrome, and $\left|\Psi_0\right>$ is the error-free state of the data qubits, while $\left|1_0\right>$ indicates a non-trivial error in the state $\left|\Psi_1\right>$ of the data qubits.  In the standard gate model, we would measure the ancilla and apply a correction to the data if the result $\left|1_0\right>$ were obtained.  In GSQC, we instead amplify the $\left|0_0\right>$ part of the wavefunction and never apply a correction.  To achieve this, we supplement the Hilbert space of the ancilla qubit with an $\left|\text{IDLE}\right>$ state and add a term to the Hamiltonian of the form $\epsilon (\Lambda \left|0_0\right> - \left|\text{IDLE}\right>) (\Lambda \left<0_0\right| - \left<\text{IDLE}\right|)/(1+\Lambda^2)$ or $\epsilon (\Lambda c^\dagger_{Q,0,0} -  d^\dagger_Q)(\Lambda c_{Q,0,0} -  d_Q)/(1+\Lambda^2)$ in second quantized notation.    Then, the new zero-energy ground state is $[(\left|0_0\right>+\Lambda\left|\text{IDLE}\right>)  \left|\Psi_0\right> + \left|1_0\right> \left|\Psi_1\right>]/(1+\Lambda^2)$.  The "no error" result $\left|\Psi_0\right>$ has now been amplified; it can be made very large by increasing $\Lambda$.

\begin{figure*}
[ptb]
\begin{center}
\includegraphics[
height=3.8207in
]%
{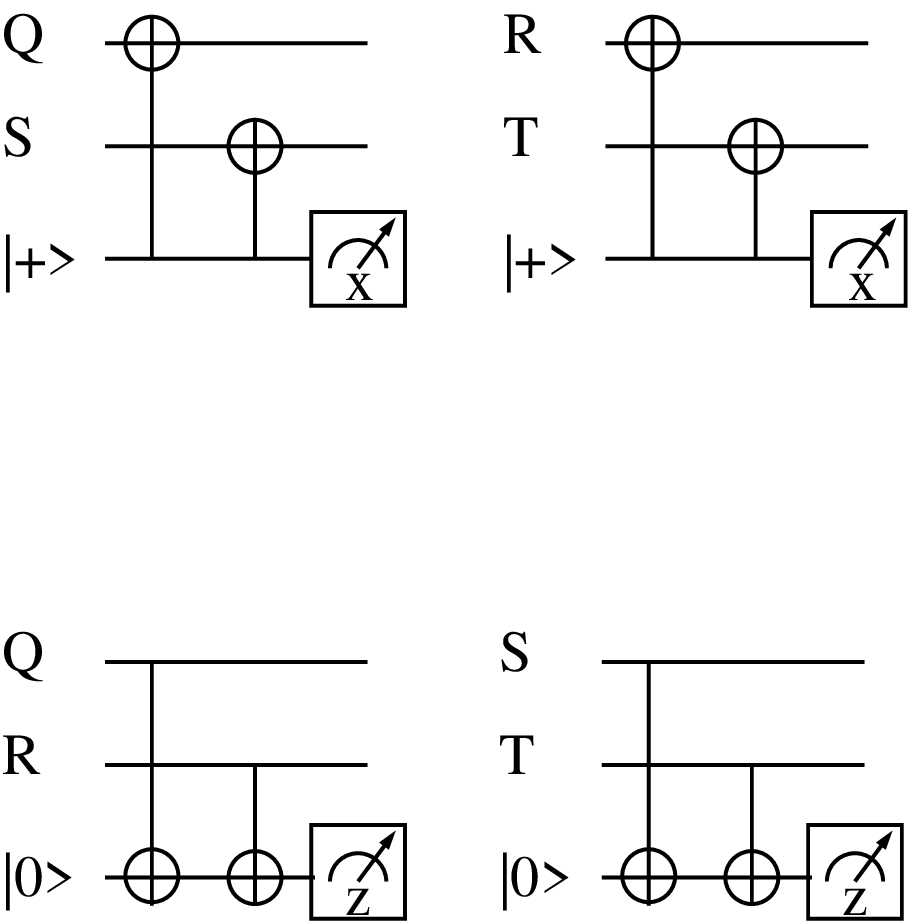}%
\caption{Error detection circuits for Bacon-Shor [[4,1,2]] code from \cite{Aliferis07b}.  The four physical qubits are labelled $Q$, $R$, $S$, and $T$.}%
\label{fig:errorcircuit}%
\end{center}
\end{figure*}

\begin{figure*}
[ptb]
\begin{center}
\includegraphics[
height=3.8207in
]%
{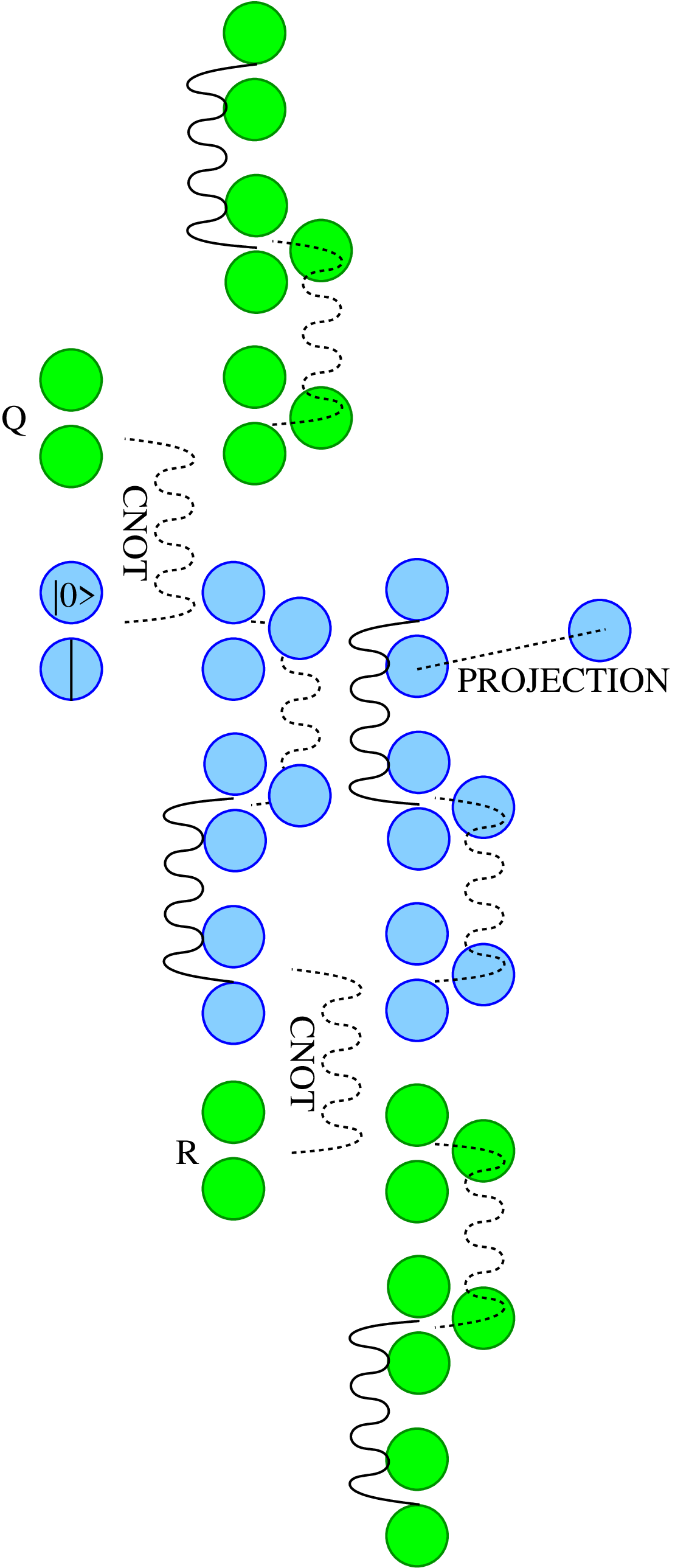}%
\caption{GSQC version of lower left error detection circuit in Fig~\ref{fig:errorcircuit}.}%
\label{fig:GSQCerror}%
\end{center}
\end{figure*}

We apply this projection method to a GSQC version of the smallest Bacon-Shor error detecting code.  This is a distance 2 subsystem code that encodes one logical qubit into four physical qubits (i.e. it is a [[4,1,2]] code).  Usually, the four physical qubits are imagined to lie in a square with coordinates $(1,1)$, $(1,2)$, $(2,1)$, and $(2,2)$, but we will label them $Q$, $R$, $S$, and $T$ respectively.  In the standard gate model, the stabilizer generators can be measured \cite{Aliferis07b} via the circuits shown in Fig.~\ref{fig:errorcircuit}.  It is straightforward to translate the CNOT gates appearing in these circuits into GSQC; this was described above.  We translate the $Z$ measurements into Hamiltonian terms $\epsilon (\Lambda \left|0_0\right> - \left|\text{IDLE}\right>) (\Lambda \left<0_0\right| - \left<\text{IDLE}\right|)/(1+\Lambda^2)$ as described above and the $X$ measurements into Hamiltonian terms $\epsilon (\Lambda (\left|0_0\right>+\left|1_0\right>)/\sqrt{2} - \left|\text{IDLE}\right>) (\Lambda (\left<0_0\right|+\left<1_0\right|)/\sqrt{2} - \left<\text{IDLE}\right|)/(1+\Lambda^2)$.  This projection amplifies the no-error portion of the wavefunction, obviating the need for a correction circuit.  A schematic of the GSQC version of the lower left circuit in Fig.~\ref{fig:errorcircuit} is shown in FIg.~\ref{fig:GSQCerror}.  To make a rough underestimate of the pseudo-threshold, we notice that there are 12 gates in the circuits in Fig.~\ref{fig:errorcircuit}, including measurements since they are translated into imperfect projections.  A CNOT extended rectangle will therefore have $4 \times 12$ error detection gates, and $4$ transverse CNOT gates.  This is a total of $52$ gates.  If $2$ gates fail, the error is undetectable and cannot be projected out, and there are about $52 \choose 2$, or about $1000$ ways for this to occur.  The resulting pseudo-threshold estimate is $1/1000 = .1\%$.  Note that (i) this pseudo-threshold is quite high compared to most in the standard gate model literature and could be an underestimate by an order of magnitude given that we didn't distinguish between malignant and benign faults \footnote{For instance, the same method underestimates the [[9,1,3]] Bacon-Shor code threshold in the standard gate model by a factor of $\sim 6$.} , (ii) the overhead involved in our error correction is exceedingly low, requiring very few extra qubits and no classical computation abilities in stark contrast to most high-threshold standard gate model error correction schemes, and (iii) this subsystem code is well-suited to locality constraints.  We considered the [[4,1,2]] code for simplicity; it is very likely that qualitative improvements will be possible by considering other existing codes and/or refining existing codes to take advantage of the properties of GSQC.

We emphasize that this error correction method will handle fabrication errors as well as the errors introduced by teleportation as long as the net probability of error is below threshold.  On the other hand, time-dependent perturbations of the system are opposed by the energy gap without any active error-correction.  A time-dependent perturbation of the form $H_{pert} = V_{pert} \cos \omega t$ can introduce an oscillating contribution $c_k(t) \left|\Psi_k\right>$ from the excited state $\left|\Psi_k\right>$ into the ground state wavefunction.  Time-dependent perturbation theory gives
\[
c_k(t) \approx - \frac{\left<\Psi_k\right|V_{pert}\left|\Psi\right>}{2\hbar} \left[\frac{e^{i(E_k/\hbar + \omega)t} - 1}{E_k/\hbar + \omega} + \frac{e^{i(E_k/\hbar - \omega)t} - 1}{E_k/\hbar - \omega} \right].
\]
If the excitation frequency is far below the gap frequency, this coefficient will not become substantial, and the measured GSQC wavefunction will not suffer distortions.  For excitations with frequencies approaching the gap frequency, transitions out of the ground state can occur at a rate approximated by Fermi's golden rule
\[
R \approx \frac{2\pi}{\hbar} \left|\left<\Psi_k\right|V_{pert}\left|\Psi\right>\right|^2 \rho(\omega).
\]

\noindent {\em Adiabatic calculation}

The adiabatic condition states that system in the ground state of the Hamiltonian will remain in the ground state if all of the parameters in the Hamiltonian change sufficiently slowly.  There have been some recent attempts to make rigorous adiabatic conditions, especially in the context of adiabatic quantum computing, but equation (\ref{eq:adiabaticcondition}) in the main text gives the traditional condition for our problem:
\begin{equation}
T \gg \hbar \Lambda_{max} \left| \left<\Psi_k\right| d{\mathcal H}/d\Lambda \left| \Psi \right>\right|/E_{k}^2 
\end{equation}
where $T$ is the time to take $\Lambda/\Lambda_{max}$ from 0 to $1$, $\left|\Psi \right> = \left|\Psi (\Lambda)\right>$ is the ground state of the Hamiltonian, $k$ labels any excited eigenstate of the Hamiltonian, and we have simplified the denominator using the fact that the energy of $\left|\Psi\right>$ vanishes.  To evaluate the right hand side, we can use the identity
\[
\frac{d^2 E_{\text{ground}}}{d\Lambda^2} = - 2 \sum_{k \ne \text{ground}} \frac{\left| \left<\Psi_k\right| d{\mathcal H}/d\Lambda \left| \Psi \right>\right|^2}{E_{k}} + \left<\Psi\right|\frac{d^2 H}{d\Lambda^2}\left|\Psi\right>.
\]
Since $E_{\text{ground}} = 0$ for all $\Lambda$, this implies that for any term in the sum
\[
\frac{\left| \left<\Psi_k\right| d{\mathcal H}/d\Lambda \left| \Psi \right>\right|}{E_{k}^2} \le \sqrt{\frac{1}{2} \frac{\left<\Psi\right|\frac{d^2 H}{d\Lambda^2}\left|\Psi\right>}{E_k^3}}.
\]
Noting that 
\begin{eqnarray*}
\frac{d^2{\mathcal H}}{d\Lambda^2} & = & \epsilon \sum_{Q,s}  \left[\frac{ c_{Q_s,1,1}^{\dagger} c_{Q_{s+1},0,0}^{\dagger}- c_{Q_s,1,0}^{\dagger} c_{Q_{s+1},0,1} ^{\dagger}}{\sqrt{2}} \,\,\, d^{\dagger}_{Q_s}d^{\dagger}_{Q_{s+1}} \right]  \\
& & \hspace{0.8in}  \left[\begin{array}{cc}2 - 6 \Lambda^2& 6 \Lambda - 2\Lambda^3 \\ 6 \Lambda - 2\Lambda^3 &2 - 6 \Lambda^2\end{array}\right] / (1 + \Lambda^2)^3\\
& & \hspace{0.8in}\left[\frac{ c_{Q_s,1,1}^{\dagger} c_{Q_{s+1},0,0}^{\dagger}- c_{Q_s,1,0}^{\dagger} c_{Q_{s+1},0,1} ^{\dagger}}{\sqrt{2}} \,\,\, d^{\dagger}_{Q_s}d^{\dagger}_{Q_{s+1}} \right] ^\dagger
\end{eqnarray*}
we find that $\left<\Psi\right|\frac{d^2 H}{d\Lambda^2}\left|\Psi\right> \le 2 N \epsilon/(1+\Lambda^2).$  The result is that it is sufficient to have
\[
T \gg \hbar  \frac{\Lambda_{max}}{\sqrt{1+\Lambda_{max}^2}} \sqrt{\frac{N \epsilon}{E(\Lambda_{max})^3}}
\]
where $E(\Lambda_{max})$ is the minimum energy gap, calculated above in the energy gap section.  Note that here $N$ is the total number of gates in the circuit.   If there are many qubits undergoing gates independently, it may be advantageous to adopt a more nuanced schedule for increasing the $\Lambda$ of the gates from $0$ to $\Lambda_{max}$.  For instance, it may be advantageous to keep some $\Lambda$ fixed at zero temporarily or to use the strategy that appears in \cite{Mizel07}.

One intriguing theoretical problem is to determine whether an adiabatic approach to the ground state is necessary or whether a cooling process is likely to find the ground state quickly. 

\bibliographystyle{unsrt}
\bibliography{fixedgap}

\end{document}